\newcommand{\be}{\begin{equation}}
\newcommand{\ee}{\end{equation}}
\newcommand{\bea}{\begin{eqnarray}}
\newcommand{\eea}{\end{eqnarray}}
\newcommand{\nn}{\nonumber \\}
\newcommand{\p}[1]{(\ref{#1})}
\documentstyle{article}
\def\a{\alpha}\def\b{\beta}
\def\d{\delta}\def\Th{\Theta}
\def\G{\Gamma}
\def\um{{\underline m}}
\def\un{{\underline n}}
\def\ui{{\underline i}}
\def\uj{{\underline j}}

\begin{document}
\begin{flushright}
DAMTP-R-97/34 \\
hep-th/97xxxxx\\
\end{flushright}

\vspace{3truecm}
\renewcommand{\thefootnote}{\fnsymbol{footnote}}
\begin{center}
{\large\bf M-theory superalgebra from the M--5--brane}

\vspace{1cm}
Dmitri Sorokin$^1$\footnote{dsorokin@kipt.kharkov.ua}
and Paul K. Townsend$^2$\footnote{p.k.townsend@damtp.cam.ac.uk}

\vspace{0.5cm}
${}^1${\it National Science Center,
Kharkov Institute of Physics and Technology,\\
Kharkov, 310108, Ukraine}

\vspace{0.5cm}
${}^2${\it DAMTP, University of Cambridge, \\
Silver St., Cambridge, U.K.}

\vspace{1.cm}
{\bf Abstract}
\end{center}
The algebra of Noether supercharges of the M-5-brane effective action is
shown to include both 2-form and 5-form central charges. Surprisingly,
only the 5-form charge is entirely due to the Wess-Zumino term because
the `naive' algebra is the M-2-brane supertranslation algebra. The full
structure of central charges is shown to be directly related to the
projector arising in the proof of $\kappa$-symmetry of the M-5-brane
action. It is also shown to allow `mixed' M-brane configurations
preserving 1/2 supersymmetry that include the (non-marginal)
M-2-brane/M-5-brane `bound state' as a special case.

\renewcommand{\thefootnote}{\arabic{footnote}}
\setcounter{footnote}0
\newpage
\section{Introduction}

The most general supertranslation algebra in eleven--dimensional
spacetime is spanned by a Majorana spinor supercharge $Q_{\a}~ (\a
=1,\dots,32)$, the 11-momentum $P_m$, a 2-form central charge
$Z_{mn}$ and a 5-form central charge $Y_{m_1\dots m_5}$
\cite{vanholten,democracy}. The only non-vanishing (anti)commutation
relation is the anticommutator
\be
\label{1}
\{ Q_{\a}, Q_{\b} \} =(C\G^m)_{\a \b} P_m +
{1\over 2}(C\G^{mn})_{\a \b}Z_{mn}+ {1\over 5!}
(C\G ^{m_1\dots m_5})_{\a \b}Y_{m_1\dots m_5},
\ee
where $\G^m$ are (constant) Dirac matrices, $\G^{m_1\dots m_p}~ (p=2,5)$
are their antisymmetrized products, and $C$ is the (real) charge
conjugation matrix. We use here the `mostly plus' metric convention for
which the Dirac matrices are real in the Majorana representation, which we
adopt below.

This superalgebra was called the M--theory superalgebra in
\cite{democracy} because its structure reflects important aspects of what
has come to be known as M--theory. For example, setting $m=(0,\um)~
(\um=1,\dots, 10)$, the positivity of the left hand side of \p{1} implies
that $P^0$ satisfies a bound of the type
\be
\label{2}
 P^0 \ge f\big( P_\um, Z_{\um\un}, Y_{\um_1\dots \um_5};
Z_{0\um}, Y_{0\um_1\dots \um_4} \big)
\ee
for some particular function $f$. Each of the arguments of this function
is a charge carried by one of the `basic' objects of M--theory. For
example, the 10-momentum $P_\um$ is associated with the D=11 massless
superparticle \cite{bergstown}, the 2-form charge $Z_{\um\un}$ is
associated with the supermembrane \cite{bst}, or M-2-brane, and (as
verified here) both, the 2-form charge and the 5-form charge
$Y_{\um_1\dots \um_5}$, are associated with the M-theory fivebrane
\cite{5b,ag}, or M-5-brane. These `basic' objects are also represented by
classical solutions of D=11 supergravity \cite{Mwave,stelle,guven}, and the
M-5-brane was originally discovered this way.  The time components of $Z$
and $Y$ are charges carried by objects that appear in Kaluza-Klein (KK)
vacua. Specifically, $Y_{0\um_1\dots \um_4}$ is carried by the D-6-brane
of IIA superstring theory while $Z_{0\um}$ can be associated with the IIA
D-8-brane\footnote{This interpretation of the time components of the
central charges has been noticed independently by C.  Hull \cite{hull}.}.
We shall not have anything further to say here about these `KK-type'
charges, as we shall assume an uncompactified D=11 spacetime. Setting them
to zero, and making use of the fact that $C=\G^0$ in the Majorana
representation, we can rewrite \p{1} as \be \label{3} \{ Q, Q\} = P^0(1 +
\bar\Gamma) \ee where \be \label{4} \bar\Gamma =
(P^0)^{-1}\big[\G^{0\um}P_\um + {1\over2} \G^{0\um\un} Z_{\um\un} +
{1\over 5!}\G^{0\um_1\dots \um_5} Y_{\um_1\dots \um_5}\big]\, .
\ee
The bound \p{2} is now equivalent to the statement that no eigenvalue of
$\bar\Gamma^2$ can exceed unity. Those choices of central charges
for which
\be
\label{5}
\bar\Gamma^2=1
\ee
are of particular significance because ${1\over2}(1+\bar\Gamma)$
is then a projector, projecting onto the eigenspace of $\bar\Gamma$ with
eigenvalue 1. Since $\bar\Gamma$ has zero trace, the dimension of this eigenspace
is half that of the original space. This means that configurations associated
with such charge values preserve 1/2 the supersymmetry.

It is easy to see that the condition \p{5} is solved
by the following three choices, in which all charges independent of the one
given vanish:
\bea
(i)~~~~~~~P_1 &=& P^0 \nn
(ii)~~~~~ Z_{12}&=& P^0 \nn
(iii)\ Y_{12345} &=& P^0 \, . \nonumber
\eea
These charges are naturally associated with a wave in the 1--direction, a
membrane in the 12--plane and a fivebrane in the 12345--plane, respectively. Of
course, these are not the only solutions of \p{5}. There is the obvious
rotational freedom, and also the freedom to consider a combination with both
$P_1$ and $Z_{23}$ non-zero, in which case
\be
\label{6}
P^0 = \sqrt{P_1^2 + Z_{23}^2} \, .
\ee
These charges have an obvious interpretation as those of a membrane in the
23--plane boosted in the 1--direction. Discounting the freedom to rotate and
boost, the configurations (i) and (ii) are the only solutions of \p{5} with
$Y=0$.

In contrast, when $Y\ne0$ there are many more possibilities than just (iii).
For example, \p{5} is also solved by
\be
\label{7}
Z_{12}=\zeta\, , \qquad
Y_{12345} = \sqrt{(P^0)^2 - \zeta^2}\, ,
\ee
where $\zeta$ is (classically) arbitrary apart from the constraint $|\zeta|\le
P^0$.  These charges can be interpreted as those of a (non-marginal)
M-2-brane/M-5-brane bound state; the corresponding solution of D=11 supergravity
preserving 1/2 supersymmetry was given in \cite{pap}. A still more general
`mixed' M-brane solution of \p{5} is
\be
\label{8}
Z_{12}=\zeta_1\, , \qquad Z_{34} =\zeta_2\, ,
\qquad P_5=\zeta_1\zeta_2\, ,
\qquad Y_{12345} =
\sqrt{(P^0)^2 - \zeta_1^2 -\zeta_2^2 -\zeta_1^2\zeta_2^2}\, .
\ee
This case, which we shall discuss further at the conclusion of this paper,
presumably corresponds to a bound state of an M-5-brane with two boosted
intersecting M-2-branes, although the associated stationary solution of D=11
supergravity is not yet known. Note that if we were to `strip off' the fivebrane
covering the two membranes (i.e. put $Y=0$) we would get a boosted configuration
of two M-2-branes intersecting at a point, which preserves 1/4 supersymmetry.
Similarly, if we were to omit the momentum then we would have a static
configuration of two intersecting membranes in a fivebrane preserving 1/4
supersymmetry. What this shows is that the addition of a brane (or a wave)
to a configuration preserving some supersymmetry may {\it increase} rather
than decrease the fraction of supersymmetry preserved.

The possibility of rotating and boosting M-branes, while preserving 1/2
supersymmetry, is reflected in the Lorentz invariance of the effective
worldvolume action. One might therefore expect the possibility of `mixed'
M-branes to be reflected in the structure of the M-5-brane effective action.
We shall confirm this expectation by showing that the supertranslation algebra
of the M-5-brane acquires a central extension with charges given precisely by
the rotationally-invariant generalization of \p{8}. In the M-2-brane case the
2-form central charge arises as a consequence of the existence of a `Wess-Zumino' (WZ) term in the effective action \cite{cc}. It was assumed in \cite{democracy} that the 5-form  charge in the supertranslation algebra would have a similar
justification, and this was made more plausible by some early partial results on
the form of the WZ term \cite{aha}, but it is only now that the
$\kappa$-symmetric M-5-brane action has been found \cite{5b,ag} that the
central charge structure can be definitively ascertained.

In carrying this out we have
found a number of unexpected features that are absent from the 2-brane case.
Firstly, it is not only $Y$ that appears as  a central charge in the M-5-brane's
supertranslation algebra, but also $Z$. This is not altogether unexpected
in view of the fact that the M-5-brane can carry the 2-form charge $Z$, in
addition to $Y$. More surprising is the fact that the WZ term in the M-5-brane
action is not exclusively responsible for this charge. This is due to the fact,
which we explain, that the underlying `naive' supersymmetry algebra is not
the standard one but is rather the M-2-brane algebra, which already has
a 2-form central charge. Taking into account {\it all} sources of central charges in the
supertranslation algebra of the M-5-brane we find (not unexpectedly) that they
are just what is required for preservation of 1/2 supersymmetry.

As is well-known, preservation of 1/2 supersymmetry by an extended object is
directly related to $\kappa$-symmetry of its effective worldvolume action, so the
$\kappa$-symmetry transformations of the M-5-brane must encode the information
about the M-theory supertranslation algebra. Indeed, we shall show that the
central charge structure of the M-5-brane's supertranslation algebra can be
deduced directly from the projector ocurring in the proof of $\kappa$-symmetry.
Related observations on the connection between supersymmetry and the
$\kappa$--symmetry projector have been made previously in \cite{bk}. Also, the
central extensions in the algebra of fermionic constraints  (or supercovariant
derivatives) have been discussed in \cite{kallosh} but, for reasons explained in
\cite{izq}, there is not generally a simple relation between the constraint
algebra and the algebra of supersymmetry Noether charges considered here.

\section{Superspace preliminaries}

We begin by summarising the salient features of D=11 superspace.
Let $Z^M = (X^m, \Th^{\a})$ be the coordinates of D=11 superspace. We may
introduce a D=11 supervielbein $E_M{}^A$ as the coordinate basis components of
the frame 1-forms $E^A = (E^a,E^\a)$. In addition to the usual superspace
3-form gauge potential $C^{(3)}$ of D=11 supergravity we introduce a 6-form
gauge potential $C^{(6)}$ \cite{cl}. The gauge-invariant field strengths are
\bea
R^{(4)} &\equiv & dC^{(3)}\, , \nn
R^{(7)} &\equiv & dC^{(6)} -{1\over2} C^{(3)} R^{(4)}\, ,
\label{9}
\eea
where the exterior product of forms is understood. The on-shell D=11
supergravity constraints imply, in particular, that these field strengths take
the form
\bea
\label{10}
R^{(4)} &=&{i\over 2}E^a  E^b E^\a  E^\b (\Gamma_{ab})_{\a\b}
+{1\over {4!}}E^a  E^b E^c E^d F^{(4)}_{dcba} \nn
R^{(7)}&=&{i\over {5!}}E^{a_1}\dots E^{a_5}  E^\a  E^\b
(\Gamma_{a_1\dots a_5})_{\a\b}
 + {1\over {7!}}E^{a_1} \dots E^{a_7} F^{(7)}_{a_7\dots a_1}\, ,
\eea
where the 7-form $F^{(7)}$ is the Hodge dual of the 4-form $F^{(4)}$. One
solution to the full set of constraints is flat D=11 Minkowski spacetime with
vanishing $F^{(4)}$; we shall call this the `D=11 vacuum'. The corresponding
superspace admits the supertranslation group as a group of isometries. This
includes the supersymmetry transformations
\be
\label{11}
\delta\Theta =\epsilon\, , \qquad
\delta X^m=i\bar\epsilon\Gamma^m\Theta\ .
\ee
Note that in the D=11 vacuum we need not distinguish between frame indices `a'
and coordinate indices `m'; we shall use the coordinate indices in what follows.

In the D=11 vacuum, $E^A$ are the left-invariant 1-forms on superspace, i.e.
$E^A=(\Pi^m,~d\Theta^{\alpha})$ where
\be
\label{12}
\Pi^m=dX^m+id\bar\Theta\Gamma^m\Theta\, .
\ee
In addition, the expressions \p{10} now simplify to
\be
\label{13}
R^{4}={i\over 2}\Pi^m \Pi^n d\Theta^\a d\Theta^\b
(\Gamma_{mn})_{\a\b}\, ,
\qquad
R^{(7)}={i\over {5!}}\Pi^{m_1}\dots
\Pi^{m_5}  d\Theta^\a d\Theta^\b
(\Gamma_{m_1\dots m_5})_{\a\b}\, .
\ee
These differential forms are clearly supertranslation invariant. Also, the
4-form $R^{(4)}$ and the 7-form $R^{(7)} + (1/2) C^{(3)} R^{(4)}$ are closed.
These facts imply that the supersymmetry variations of $C^{(3)}$ and
$C^{(6)}$ take the form
\bea
\delta C^{(3)} &=& id[ \bar\epsilon \Delta_2] \nn
\delta C^{(6)} &=& id[\bar\epsilon \Delta_5] +
{i\over2} (\bar\epsilon\Delta_2) R^{(4)}\, ,
\label{14}
\eea
where the spinor-valued p-form $\Delta_p$ ($p=2,5$) takes the form
\be
\label{15}
\Delta_p ={1\over {p!}}
\Gamma_{m_1\dots m_p}\Theta\Pi^{m_1} \dots \Pi^{m_p} ~+~ \dots
\ee
where `$\dots$' indicates terms cubic or higher in $\Theta$; the full
expression for $\Delta_2$ can be found in \cite{cc}. We shall need the
supersymmetry variations of $\Delta_2$ and $\Delta_5$. These are determined
by cohomological descent to be
\bea
\delta\Delta_2 &=& {\cal A}_2 \epsilon\, ,\label{16}\\
\delta\Delta_5 &=& [{\cal A}_5 -{1\over2}{\cal A}_2 C^{(3)}]\epsilon
\label{17}
\eea
where the matrix-valued p-forms ${\cal A}_p$ have the form
\be
\label{18}
{\cal A}_p = {1\over {p!}}
\Gamma_{m_1\dots m_p}dX^{m_1}\dots dX^{m_p} ~+~ d\Lambda_{(p-1)}\, ,
\ee
The specific forms of $\Lambda_{p-1}$ are not relevant here but can be found for
$p=2$ in \cite{cc}. Note that the entries of ${\cal A}_p$ are p-forms which
are closed but not exact, in an appropriate cohomology (de Rahm in the
case of toroidally compactified space).

\section{M-5-brane: $\kappa$-symmetry and supersymmetry}

We must now review the structure of the manifestly  $d=6$ general
coordinate invariant form \cite{5b} of the M-5-brane action.

Let $\xi^i ~(i=0,1,\dots,5)$ be the worldvolume coordinates of the fivebrane. The
worldvolume fields comprise the maps $Z^M(\xi)$ from the worldvolume to
superspace, and a 2-form gauge potential
$A(\xi)$ with `modified' field strength \cite{pktb}
\be
\label{19}
H_{ijk}={\partial}_{[i}A_{jk]}-C^{(3)}_{ijk}\, ,
\ee
where $C^{(3)}_{ijk}$ is the pullback of the superspace 3-form gauge potential
$C^{(3)}$. In order that the M-5-brane action be invariant under
super-isometries of the D=11 vacuum background we must require $H$ to be
supersymmetry invariant. In view of the supersymmetry transformation of
$C^{(3)}$ we must set
\be
\label{20}
\delta A = i\bar\epsilon \Delta_2 \, ,
\ee
where $\Delta_2$ is here to be understood as the pullback to the worldvolume of
the superspace 2-form introduced in \p{14}. For the construction of the
M-5-brane action, we also need the worldvolume six-form $C^{(6)}_{i_1\dots i_6}$
induced by the superspace 6-form gauge potential, and the induced worldvolume
metric $g_{ij}(\xi )= E_i{}^a E_j{}^b \eta_{ab}$, where $\eta$ is the D=11
Minkowski metric and $E_i{}^a =\partial_i Z^M E_M{}^a$.

The 3-form field strength $H$ is required to be self-dual in a
generalized sense, the self--duality condition being a consequence of the
gauge field $A$ equation of motion \cite{ps}. The manifestly $d=6$ general
coordinate invariance of the M-5-brane action is achieved if we introduce
an auxiliary worldvolume scalar field $a(\xi)$ \cite{pst} (which is inert
under super-isometries of the D=11 superspace background). Defining
$g=\det(g_{ij})$, and \be \label{21} (H^*)^{ijk} = {1\over 3!\sqrt{-g}}
\varepsilon^{ijki'j'k'}H_{i'j'k'} \, ,\qquad {\tilde H}^{ij} = {1\over
\sqrt{-(\partial a \cdot \partial a)}}\, (H^*)^{ijk}\partial_k a\, , \ee
we can now write the M-5-brane action as
\be
\label{22}
S= \int\! d^6\xi\, (L_0 + L_{WZ})\, ,
\ee
where
\bea
\label{23}
L_0 &=& -\sqrt{-\det(g_{ij}+{\tilde H}_{ij})}
+ {\sqrt{-g} \over 4(\partial a \cdot \partial a)}
({\partial}_i a) (H^*)^{ijk} H_{jkl}({\partial}^l a) \label{23a}\\
L_{WZ} &=& {1\over{6!}}\varepsilon ^{i_1\dots i_6}
\big[C^{(6)}_{i_1\dots i_6}+
10 H_{i_1i_2i_3} C^{(3)}_{i_4i_5i_6}\big] \, .
\eea

The key feature of this action is its invariance under $\kappa$-symmetry
transformations. On general grounds the variation $\delta_\kappa Z^M E_M{}^\a$
must take the form $[{\cal P}\kappa]^\a$ where $\kappa(\xi)$ is the D=11 spinor
parameter and ${\cal P}$ is a projector with ${\rm tr} {\cal P} =16$. Such a
projector can be written as ${\cal P}= (1/2)(1+ \G_*)$ where the matrix $\G_*$
is  tracefree and squares to the identity. For the M-5-brane we have \cite{5b}
\bea
\G_* =&& {1 \over \sqrt{\det(g_{ij}+\tilde H_{ij})}}
\bigg[(\partial_i a \G^i)\G_jt^j -
{\sqrt{-g}\over 2\sqrt{-(\partial a \cdot \partial a)}}\,
(\partial_i a \G^i)\G^{jk} \tilde H_{jk} \nn
&& +\qquad {1\over 5! (\partial a \cdot \partial a) }\,
(\partial_i a \G^i) \G_{i_1\dots i_5}\,
{\varepsilon}^{i_1\dots i_5 j} (\partial_j a) \bigg]\, ,
\label{24}
\eea
where (note that $t^i{\partial}_i a \equiv 0$)
\be
\label{25}
t^i = {1\over 8(\partial a\cdot \partial a)}\,
{\varepsilon}^{ij_1j_2k_1k_2\ell}
{\tilde H}_{j_1j_2}{\tilde H}_{k_3k_4} \partial_\ell a
\ee
and ${\G}_i=E_i{}^a\G_a$ are the pullbacks to the worldvolume of the $D=11$ Dirac
matrices.

The matrix $\G_*$ has similar properties to the matrix $\bar\G$ introduced
previously in the discussion of the supertranslation algebra. It also has a
similar structure, which is even more evident if in
\p{24} we choose the temporal gauge
\be
\label {26}
a(\xi) =t\, ,
\ee
which is possible because of the invariance of the M-5-brane action under the
local transformations $\d a=\phi(\xi ),~~ \d A={\phi}(\xi )f$,
where $f$ is a worldvolume 2-form constructed from $H$ and $a$ and given
explicitly in \cite{5b}. In this gauge, and considering an infinite planar
fivebrane in the D=11 vacuum, for which the induced metric is flat, we find that
\be
\label{27}
\G_* =
{1\over \sqrt{\det(\d_{ij}+ \tilde H_{ij})}}
[\G^0 \G_\ui t^\ui - {1\over2}\G^0 \G^{ij}{\tilde H}_{ij}
+{1\over{5!}}\G^0 \G_{\ui_1,\dots, \ui_5} {\varepsilon}^{\ui_1\dots \ui_5}]\, .
\ee
We may choose five of the ten space coordinates $X^\um$ to be the space
coordinates $\sigma^\ui~ (\ui=1,\dots,5)$ of the fivebrane. If we then set
\be
\label{28}
\tilde H_{ij}= - Z_{ij}
\ee
and take the time component of $Z$ to vanish, as before, we find that
$\G_*=\bar\G$ with
\bea
Y^{\ui_1\dots \ui_5} &=& \varepsilon^{\ui_1\dots \ui_5}\nn
P^\ui &=& {1\over8} \varepsilon^{\ui\, \uj_1 \uj_2 \uj_3 \uj_4} \,
Z_{\uj_1\uj_2}Z_{\uj_3\uj_4}\nn
P^0 &=& \sqrt{ \det (\d_{\ui\,\uj} + Z_{\ui\,\uj})}\, ,
\label{29}
\eea
all other components of the charges vanishing. The construction guarantees that
this charge configuration preserves 1/2 supersymmetry but this can be verified
directly by use the following identity satisfied by any
antisymmetric $6\times 6$ matrix $Z$:
\be
\label{30}
\det(1+Z)= 1-{1\over 2}tr{Z}^2+{1\over 8}(tr{Z}^2)^2-
{1\over 4}tr{Z}^4\, .
\ee
By means of an $SO(5)$ transformation we can bring the 2-form charge $Z$ to a
form in which it has only two independent non-zero components
$Z_{12}=-Z_{21}=\zeta_1$, $Z_{34}=-Z_{43}=\zeta_2$. This yields the charge
configuration of \p{8}. Thus, what we have found here from $\kappa$-symmetry
considerations is the $SO(5)$ invariant generalization of the charge
configuration \p{8}, derived  there from the requirement of preservation of 1/2
supersymmetry.

In view of the well-known connection between the preservation of 1/2
supersymmetry by extended objects and $\kappa$-symmetry of their effective
actions, it is not too surprising that the $\kappa$-symmetry projector encodes
the form of the central charges in the M-5-brane superalgebra that ensures
preservation of 1/2 supersymmetry. As things stand, however, this connection is
no more than an observation that two matrices happen to coincide if the
variables in one are related to those in the other in a particular way. To show
that this is no mere coincidence we must compute the central charge structure of
the M-5-brane superalgebra and verify that the charges are given by \p{28} and
\p{29}.

\section{M-5-brane superalgebra}

We have seen that the M-5-brane Lagrangian $L$ can be written as
$L=L_0+L_{WZ}$. In the D=11 vacuum the Lagrangian $L_0$ is
invariant under the global supersymmetry transformations
\be
\label{31}
\delta\Theta =\epsilon\, , \qquad
\delta X^m=i\bar\epsilon\Gamma^m\Theta\, , \qquad \delta A = i\bar\epsilon
\Delta_2\, .
\ee
The anticommutators of the corresponding Noether charges, computed via the
canonical (anti)commutation relations of the worldvolume fields, yield what we
will call the `naive' supertranslation algebra. The form of this algebra is not
specific to $L_0$; we would get the same result for any Lagrangian invariant
under the supersymmetry transformations \p{31}. The reason for the terminology
`naive' is that the true supersymmetry algebra of the M-5-brane will be a
central extension of the naive one as a consequence of the fact that the WZ
term $L_{WZ}$ is not invariant under the transformations \p{31} but rather, as
we shall see, changes by a total derivative.

One might suppose that the `naive' supersymmetry algebra is just the standard
one, as is the case for the usual form of the $\kappa$-symmetric
M-2-brane action. But this is not so for the M-5-brane. The commutator of two
supersymmetry transformations acting on $A$ is
\be
\label{32}
\{Q,Q\}A={\cal A}_2 \, ,
\ee
where ${\cal A}_2$ (eq. \p{18}) is now to be understood as the pullback to
the world volume of the matrix valued 2-form found from the supersymmetry
variation of $\Delta_2$.  Thus,
\be \label{33} \{Q,Q\}A_{ij} = \partial_iX^m\partial_jX^n\Gamma_{mn}
+ \partial_{[i}\Lambda_{j]}.
\ee
When this is
integrated over a 2-cycle ${\cal M}_2$ in the fivebrane we find \be
\label{34}
\{Q,Q\}\int_{{\cal M}_2} A = {1\over2}\Gamma_{mn}\int_{{\cal M}_2}dX^m dX^n\, ,
\ee
but the right hand side is just the 2-form central charge occurring in the
M-2-brane supertranslation algebra.

In the Hamiltonian formulation the supersymmetry charge $Q^{(0)}_\a$
which generates \p{31}--\p{34} is expressed as an integral over the
fivebrane at fixed time, ${\cal M}_5$, as follows
\be
\label{35}
Q^{(0)}_\a = i\int\! d^5\sigma \big[ (\pi + i\Theta \G^m {\cal P}_m) +
i{\cal P}^{\ui\,\uj}(\Delta_2)_{\ui\,\uj}\big]\ ,
\ee
where $\pi$, ${\cal P}_m$ and ${\cal P}^{\ui\,\uj}$ are the variables
canonically conjugate to $\Theta$, $X^m$ and $A_{\ui\,\uj}$, respectively,
derived from the full M-5-brane Lagrangian $L_0+L_{wz}$ (we shall give the
explicit form of ${\cal P}_m$ and ${\cal P}^{\ui\,\uj}$ below).
Using the canonical quantum (anti)commutation relations we find that
\be
\label{37}
\{Q^{(0)}_\a,Q^{(0)}_\b\} = (C\G^m)_{\a\b}P_m  +
{1\over2} (C\G_{mn})_{\a\b}Z_0^{mn}\, ,
\ee
where $P_m$ is the integral over ${\cal M}_5$ of the density ${\cal P}_m$, and
\be
\label{38}
Z_0^{mn} = -\int_{{\cal M}_5} dX^m dX^n{\cal P}^* \, ,
\ee
 with ${\cal P}^*$ the 3-form dual of ${\cal P}$, i.e.
${\cal P}^*_{\ui_1\ui_2\ui_3} =
{1\over2}\varepsilon_{\ui_1\ui_2\ui_3\ui_4\ui_5}{\cal P}^{\ui_4\ui_5}$.

We conclude that the `naive' supertranslation algebra, i.e. the algebra of
Noether charges that would be associated with an invariant fivebrane Lagrangian,
already includes a 2-form central charge!
In fact, there is a sense in which this is already true for the M-2-brane.
In the `scale-invariant' formulation of the M-2-brane action \cite{blt}
the WZ term is replaced by a two-form gauge potential with essentially the
same `modified' field strength tensor as that of the M-5-brane 2-form field
$A$ (the only difference is the dimension of the worldvolume on which the
3-form field strength is defined). In this formulation of the M-2-brane
the `naive' algebra is a 2-form central extension of the standard
supertranslation algebra. This leads us to expect that the supertranslation
algebra of the complete M-5-brane action will also contain a 2-form charge
proportional to the constant `expectation value' of the 3-form $H$. As we
shall see below there are actually two equal contributions of this type;
one is the `naive' contribution under discussion here while the other
arises from the non-invariance of the WZ term.

So we turn now to the WZ Lagrangian. It is convenient to rewrite it in
differential form notation as
\be
\label{39}
L_{WZ} = C^{(6)} + {1\over2} H\wedge C^{(3)}\, .
\ee
Its supersymmetry variation is
\be
\label{40}
\delta L_{WZ} = id(\bar\epsilon \Delta)\ \qquad
(\Delta \equiv \Delta_5 - {1\over 2}\Delta_2 H)\, ,
\ee
where the p-forms $\Delta_p$  are to be understood as
pullbacks to the worldvolume of the corresponding superspace p-forms
defined in \p{16}, so $\Delta$ is a worldvolume 5-form. We see that the
supersymmetry variation of the M-5-brane action includes a boundary
term. This term contributes to the supercharge, which has the form
\bea
Q_\alpha &=& Q^{(0)}_\alpha +\int_{{\cal M}_5}\! \Delta_\alpha \nn
&=& i\int\! d^5\sigma\,\big[(\pi+i\bar\Theta\Gamma^{m}{\cal P}_{m})_\alpha
+i({\cal P}^{\ui_1\ui_2} + {1\over 4}
{H}^{*0\ui_1\ui_2})(\Delta^2_{{\ui_1\ui_2}})_\alpha
-i\varepsilon^{\ui_1...\ui_5}
(\Delta^5_{\ui_1...\ui_5})_\alpha\big]\, .
\label{q}
\eea
Taking the (quantum) anticommutator of these supercharges we find that
\be\label{ma}
\{Q_\a,Q_\b\} = (C\G^m)_{\a\b}P_m + {1\over 2} (C\G_{mn})_{\a\b}
Z^{mn}+{1\over{5!}}\Gamma_{m_1\dots m_5}Y^{m_1\dots m_5}\, ,
\ee
where
\be\label{five}
Y^{m_1\dots m_5} = \int_{{\cal M}_5} dX^{m_1} \cdots dX^{m_5}\, ,
\label{42}
\ee
while the 2-form central charge is now the sum $Z=Z_0 + Z_{\scriptscriptstyle
WZ}$, where $Z_0$ is given by \p{38} and
\be
\label{WZpart}
Z_{\scriptscriptstyle WZ}^{mn} = -{1\over2}
\int_{{\cal M}_5} dX^m dX^n\, (H+C^{(3)}) \, .
\ee
The $C^{(3)}$ term in this expression is due to the $C^{(3)}$ term in \p{17}.

To
compute $Z_0$ we choose the temporal gauge \p{26}, then:
\be \label{pab} {\cal P}^{\ui\,\uj} \equiv {1\over{\sqrt{-g}}}{{\delta
L}\over{\delta(\partial_0A_{\ui\,\uj})}} = {1\over
4}(H^{*0\ui\,\uj}+C^{*0\ui\,\uj})\, .
\ee
Equivalently ${\cal P}^* =
(1/2)(H+C^{(3)})$, so $Z_0=Z_{\scriptscriptstyle WZ}$. Since
$H+C^{(3)}=dA$ we conclude\footnote{It also follows that the expression
\p{pab} for ${\cal P}$ does not contain time derivatives and is therefore
a constraint. This constraint reflects the self--duality of the
worldvolume field strength $H$.} that
\be \label{topo} Z^{mn} =
-\int_{{\cal M}_5} dX^m dX^n\, dA\, .
\ee
Like $Y$ (and in accord with
general principles \cite{izq}), this is a topological charge. It is
non-zero only for topologically nontrivial configurations of the
self--dual field; for instance, when $dA$ is a constant 3-form. The 5-form
charge $Y$ is just the electric source of $C^{(6)}$ or, equivalently, the
magnetic source of $C^{(3)}$. The 2-form charge $Z$ is the electric source
of $C^{(3)}$, as is easily seen by variation of the M-5-brane action with
respect to this background field which couples to the fivebrane via the
`modified' 3-form field strength $H$.  What we have now shown is that both
these M-brane charges appear as central charges in the M-5-brane
supertranslation algebra.

If we now consider an infinite planar fivebrane in
the D=11 vacuum and choose five of the ten space coordinates $X^\um$ to coincide
with the five space coordinates $\sigma^\ui$ of the fivebrane then the central
charges $Z$ and $Y$ can be written as
\be
\label{44}
Y^{\ui_1\dots \ui_5} = \varepsilon^{\ui_1\dots \ui_5}\, \qquad
Z^{\ui\uj} = -{\tilde H}^{\ui\uj}\, ,
\ee
with all other components vanishing, and where $\tilde H$ is now to be
understood as a constant `expectation value' of the worldvolume field.
Note that $C^{(3)}=0$ and the induced metric is flat in this
case, so $\tilde H_{\ui}{}^{\uj} = \tilde H_{\ui\uj}$.

To complete the determination of the $\{Q,Q\}$ anticommutator we must
compute the `non-anomalous' term proportional to $P_m$. In the context of
the M-5-brane action $P_m$ is just the integral over the fivebrane of the
variable conjugate to $X^m$:
\be
\label{45}
P_{m}=\int d^5\sigma\, {\delta L\over \delta(\partial_t X^m)}\, .
\ee
Explicit computation leads to the conclusion that the only non-zero
components of $P_m$ are $P_0$ and $P_\ui$ (the components of $P_\um$
parallel to the fivebrane). Using the relation in \p{44} between $H$ and the
2-form charge $Z$ we then find (for an infinite planar fivebrane in the D=11
vacuum with a flat induced metric)
\be
\label{46}
P^0 = \sqrt{\det (\d_{\ui\,\uj} + Z_{\ui\,\uj})}\, , \qquad
P^\ui = {1\over8}\varepsilon^{\ui\,\uj_1\uj_2\uj_3\uj_4}
Z_{\uj_1\uj_2}Z_{\uj_3\uj_4}\, .
\ee
Of course, $P^0$ must be interpreted as the fivebrane tension.
The fact that the membrane charge contributes to the M-5-brane tension is
not surprising. The fact that the momentum is generally forced to be non-zero
is somewhat surprising; it is a consequence of the term in $L_0$
quadratic in $H$.

We have now determined the full supertranslation algebra of the M-5-brane.
It has a 2-form and a 5-form central charge given by \p{topo}, \p{42}
and the 11-momentum given by \p{46}. These are precisely the results of \p{28}
and \p{29} anticipated earlier by consideration of $\kappa$-symmetry; as we
saw there, these charges are just such as to ensure the preservation of
1/2 supersymmetry.

\section{Comments}

We have seen that the M-5-brane supertranslation algebra allows the
possibility of `mixed' M-brane configurations preserving 1/2
supersymmetry. In the absence of KK-branes the time components of the
2-form $Z$ and the 5-form $Y$ vanish.  The spatial components of these
charges in directions orthogonal to the fivebrane also vanish. The remaining
non-zero components can be brought, by an $SO(5)$ transformation, to the form
\p{8} in which $\zeta_1$ and $\zeta_2$ can be interpreted as the charges
associated with two overlapping M-2-branes stretched along orthogonal
directions inside the fivebrane. The whole configuration of branes moves
along the fifth direction of the fivebrane with the momentum
$\zeta_1\zeta_2$ (or one can say that a wave propagates along the fivebrane
in this direction). A slight generalization is possible (still preserving
1/2 supersymmetry) in  which $P$ has a non-zero component orthogonal to
the fivebrane, but this corresponds to a boost of the configuration just
described.

For the special case in which $\zeta_2=0$ the associated 1/2
supersymmetric solution of D=11 supergravity is known \cite{pap}.  The
D=11 supergravity solution corresponding to the more general case, which
will be stationary rather than static, is not yet known. It seems likely
that it could be constructed as the lift to D=11 of a U-dualized extreme
black hole solution in D=6. Given such a solution, it could be
dimensionally reduced to a  static solution of IIA supergravity, in which
context it could be interpreted as a non-marginal bound state in IIA
superstring theory of a D-0-brane at the intersection of two D-2-branes
within a D-4--brane. This type of `mixed' D-brane configuration has been
discussed previously by various authors, e.g. \cite{co}. These
possibilities could of course be deduced directly from the central charge
structure of the D-p-brane supertranslation algebras, but the results so
obtained must be related by dualities to those obtained here for the
M-5-brane.

Returning to the M-theory algebra in the form \p{3} we can ask whether
there are any further possibilities with $\bar \Gamma^2=1$ that could have
a `mixed' M-brane interpretation. We have found only one example
(excluding KK charges). This is when $P$, $Z$ and $Y$ each have spatial
components that are completely orthogonal to each other (for instance,
$P_1$, $Z_{23}$ and $Y_{45678}$). Since, in this case, non-zero momentum
corresponds to an orthogonal boost we can set it to zero without loss of
generality. The resulting $Z$ and $Y$ charges could, in principle,
correspond to a non-marginal bound state of an M-2-brane orthogonally
intersecting an M-5-brane. If there were such a bound state then duality
would imply the existence of a non-marginal 0-brane/6-brane bound state
(preserving 1/2 supersymmetry) in IIA superstring theory, but no such
bound state exists because the force between the constituents is repulsive
rather than attractive \cite{pol}. We thus conclude that {\sl all
non-marginal bound states of M-branes preserving 1/2 supersymmetry are
accounted for by the M-5-brane effective action}. Since M-wave and
M-2-brane are in the same equivalence class as the M-5-brane under duality
(at least after compactification on $T^2$) we see that M-theory is
essentially the theory of a single object, but one which takes on various
forms in various dual formulations of the theory.

One surprising result of our analysis is that the 2-form central extension
in the M-5-brane algebra is not entirely due to the WZ term in the action.
The source of the other contribution is reminiscent of the source of the
2-form central charge in the `scale-invariant' formulation \cite{blt} of
the M-2-brane action. Together, these facts suggest that the realization
of supersymmetry as translations in superspace is not the ideal way to
think about it. Instead, one needs to consider something else, perhaps a
free-differential algebra, in which the ocurrence of the 2-form charge in
the supertranslation algebra is automatic.

\bigskip
\noindent
{\bf Acknowledgements.} The authors are thankful to I. Bandos, E.
Bergshoeff, V. Balasubramanian, M. Costa, R. Kallosh, J. Maldacena, M. de
Roo, D.  Polyakov, M. Tonin, A. Tseytlin and  M.  Vasiliev for useful
discussion.  Work of D.S. was partially supported by research grants of
the Ministry of Science and Technology of Ukraine and the
INTAS Grants N 93--127--ext. and N 93--493--ext.


\bigskip
\bigskip
\begin{thebibliography}{99}

\bibitem{vanholten}
J.W. van Holten and A. Van Proeyen, {\sl J. Phys. A: Math Gen.} {\bf 15}
(1982) 3763.

\bibitem{democracy}
P.K. Townsend, {\it p-brane democracy}, hep-th/9507048.

\bibitem{bergstown}
E. Bergshoeff and P. K. Townsend, {\sl Nucl. Phys.}
{\bf B490} (1997) 145.

\bibitem{bst}
E. Bergshoeff, E. Sezgin and P. K. Townsend, {\sl Phys. Lett.} {\bf
189B} (1987) 75; {\sl Ann. Phys. (N.Y.)} {\bf 185} (1988) 330.

\bibitem{5b}
P. Pasti, D. Sorokin and M. Tonin, {\sl Phys. Lett.} {\bf 398B}
(1997) 41;\\
I. Bandos, K. Lechner, A. Nurmagambetov, P. Pasti, D. Sorokin and M.
Tonin, {\sl Phys. Rev. Lett.} {\bf 78} (1997) 4332.

\bibitem{ag}
M. Aganagic, J. Park, C. Popescu and J. H. Schwarz, {\sl Nucl. Phys.}
{\bf B496} (1997) 191.

\bibitem{Mwave}
C.M. Hull, Phys. Lett. {\bf 139B} (1984) 39.

\bibitem{stelle}
M. J. Duff and K. S. Stelle, {\sl Phys. Lett.} {\bf 253B} (1991) 113.

\bibitem{guven}
R. G\"uven, {\sl Phys. Lett.} {\bf 276B} (1992) 49.

\bibitem{hull}
C.M. Hull, {\sl Gravitational duality, branes and charges}, hep-th/9705162.

\bibitem{pap}
J. M. Izquierdo, N. D. Lambert, G. Papadopulos and P. K. Townsend,
{\sl Nucl. Phys.} {\bf B460} (1996) 560.

\bibitem{cc}
J. A. de Azcarraga, J. P. Gauntlett, J. M. Izquierdo and P. K. Townsend,
{\sl Phys. Rev. Lett.} {\bf 63} (1989) 2443.

\bibitem{aha}
O. Aharony, {\sl Nucl. Phys.} {\bf B476} (1996) 470;\\
E. Bergshoeff, M. de Roo and T. Ort\'\i n, {\sl Phys. Lett.} {\bf 386B}
(1996) 85.

\bibitem{bk}
E. Bergshoeff, R. Kallosh, T. Ort\'\i n and G. Papadopulos, {\it
$\kappa$-symmetry, Supersymmetry and Intersecting Branes}, hep--th/9705040.

\bibitem{kallosh}
R. Kallosh, {\it Covariant quantization of D--branes}, hep--th/9705056.

\bibitem{izq}
J.A. de Azc{\' a}rraga, J.M. Izquierdo and P.K. Townsend, {\sl Phys. Lett.} {\bf
267B} (1991) 366.

\bibitem{cl}
A. Candiello and K. Lechner, {\sl Nucl. Phys.} {\bf B412} (1994) 479;\\
P.S. Howe, {\it Weyl Superspace}, hep-th/9707184.

\bibitem{pktb}
P.K. Townsend, {\sl Phys. Lett.} {\bf B373} (1996) 68.

\bibitem{ps}
 M. Perry and J. H. Schwarz, {\sl Nucl. Phys.} {\bf 498} (1997) 47,
hep-th/9611065.

\bibitem{pst}
P. Pasti, D. Sorokin and M. Tonin, {\sl Phys. Rev.} {\bf D52} (1995) R4277;
{\it Ibid} {\bf D55} (1997) 6292.

\bibitem{blt}
E. Bergshoeff, L.A.J. London and P.K. Townsend,  Class. Quantum Grav. {\bf 9}
(1992) 2545;\\
P.K. Townsend, {\sl Membrane tension and manifest IIB S-duality},
hep-th/9705160.

\bibitem{co}
M. S. Costa and M. Cveti$\tilde{\rm c}$, {\it Non--threshold D--brane bound
states and black holes with non--zero entropy}, hep-th/9703204;\\
A. A. Tseytlin, {\it Composite BPS configurations of p--branes in 10 and
11 dimensions}, hep--th/9702103.

\bibitem{pol}
J. Polchinski, {\it Tasi lectures on D-branes}, hep-th/9611050.

\end{thebibliography}
\end{document}